# Effect of Carbon on the Compound Layer Properties of AISI H13 Tool Steel in Pulsed Plasma Nitrocarburizing

Rodrigo L. O. Basso, Carlos A. Figueroa, Luiz F. Zagonel, Heloise O. Pastore, Daniel Wisnivesky, Fernando Alvarez*

Due to the mechanical and inertness properties of the $\varepsilon$-Fe$_{2\text{-}3}$N phase, its formation as a compact monolayer is most wanted in plasma surface treatments of steels. This phase can be obtained by the inclusion of carbon species in the plasma. In this work, we present a systematic study of the carbon influence on the compound layer in an AISI H13 tool steel by pulsed plasma nitrocarburizing process with different gaseous ratios (0% $\leq$ [CH$_4$]/ [N$_2$ + CH$_4$ + H$_2$] $\leq$ 4%). The plasma treatment was carried out for 5 h at 575 °C. The microstructure and phase composition of the modified layers were studied by scanning electron microscopy and X-ray diffraction, respectively. X-Ray photoelectron spectroscopy was used to measure the relative concentration of carbon and nitrogen on the surface. The hardening profile induced by the nitrocarburized process is also reported.

## Introduction

The mechanical and inertness properties of a compact compound monolayer, constituted of the $\varepsilon$-Fe$_{2\text{-}3}$(C,N) phase, are pursued in some applications where hardness and corrosion resistance are wanted in metals.[1–6] The use of plasma nitriding in steel surface modifications is a well-established method since the compound and diffusion layers are generally well controlled by the technique. On the other hand, to obtain a compact, monolayer $\varepsilon$-Fe$_{2\text{-}3}$(C,N) phase by plasma nitriding is not straightforward. Therefore, in order to overcome this inconvenience, the nitrocarburizing process proved to be a possible fashion to obtain a monolayer formed by the $\varepsilon$-Fe$_{2\text{-}3}$(N,C) phase.[7,8] In general, plasma nitrocarburizing normally produces a compound layer with mixed $\varepsilon$-Fe$_{2\text{-}3}$(N,C), $\gamma'$-Fe$_4$N and/or $\theta$-Fe$_3$C phases. Such a nitrocarburized phase structure is known to be deleterious to the tribological properties due to the inherently high internal stresses arising from the phase boundaries. This stress is due, among other things, to the difference in the lattice parameters of the phases constituting the compound layer.[3] However, the Fe-N-C ternary phase diagram proposed by Slycke et al.[9] shows a possible route leading to a single, $\varepsilon$-phase monolayer by controlling the nitrogen and carbon content on the forming layer at the surface. Indeed, for a relative low carbon concentration level in the plasma atmosphere, the undesirable $\gamma'$-phase is also formed. On the other hand, by increasing the carbon concentration in the atmosphere, a layer containing cementite is obtained which is also undesirable. As it is well known, the type of phase formation and characteristics of the surface layers are specific to the treatment, depending on the material composition and process parameters.[3] Recently, a monophase $\varepsilon$ compound layer was achieved on a plain carbon steel surface in austenitic plasma

R. L. O. Basso, C. A. Figueroa, L. F. Zagonel, D. Wisnivesky, F. Alvarez
Instituto de Física ''Gleb Wataghin'', Universidade Estadual de Campinas, Unicamp, CEP 13083-970 Campinas, SP, Brazil
E-mail: alvarez@ifi.unicamp.br
H. O. Pastore
Instituto de Química, Universidade Estadual de Campinas, Unicamp, CEP 13084-862, Campinas, SP, Brazil

nitrocarburizing by using an organic vapor as the carbon source.[10] In the present study, the purpose is investigating the optimal treatment condition by plasma nitrocarburizing of a martensitic steel leading to the formation of a compact, homogeneous $\varepsilon$-Fe$_{2-3}$(C,N) phase monolayer.

## Experimental Part

Rectangular substrates (20 × 10 × 2 mm$^3$) from commercial AISI H13 tool steel of composition 90.6Fe, 0.5C, 0.4Mn, 1.0Si, 5.1Cr, 1.4Mo, 0.9V (in wt.-%) were used for the experiments. The ferritic steel was pretreated by quenching and tempering, to obtain martensitic microstructure. Before nitrocarburizing, the samples were mirror polished with diamond abrasive powder (1 μm) and ultrasonically cleaned in alcohol. The nitrocarburizing experiments were performed in a plasma apparatus consisting of a hot wall furnace with 200 kg load capacity powered by a pulsed source of 1 000 V and 60 A maximum voltage and current, respectively.[11] A low pressure sputter precleaning was realized with Ar and H$_2$ plasma during the heating of the samples up to the nitrocarburizing temperature (575 °C ± 5). The plasma nitrocarburizing process was performed at a constant pressure of 400 Pa during 5 h. For the study, different gaseous mixtures of N$_2$, H$_2$, and CH$_4$ were used (0% ≤ $\phi$ ≤ 4%, $\phi$ = [CH$_4$]/[N$_2$ + CH$_4$ + H$_2$]). The total flux was maintained constant at 0.35 slm and the different gaseous mixtures were obtained by introducing methane at different proportions (0, 1, 2, 3, and 4%). The remaining gaseous flux, completing the 0.35 slm, was supplied by N$_2$ and the H$_2$ at the ratio of 90/10; i.e., 90% N$_2$ and 10% H$_2$. After the treatment, the samples were cooled down slowly inside the vacuum chamber. The treated samples were sliced and mounted in conductor Bakelite and mirror polished (0.05 μm) with colloidal silica. The carbonitrided layers were revealed, at room temperature, by etching the samples with Nital 2% (2% v/v nitric acid in absolute ethanol). The cross-section morphology and the thickness of the carbonitrided layers were analyzed by SEM (JEOL JMS-5900LV). In all the carbonitrided samples, the hardness was obtained using a Berkovich diamond tip (NanoTest-300) and the results analyzed using the Oliver and Pharr method.[12] Each reported hardness measurement was obtained averaging ten indentation curves. The phase compositions of the compound layers were studied by X-ray diffraction (XRD) analysis, performed in both $\theta$-2$\theta$ and glancing incidence geometry (GIXRD). The $\theta$-2$\theta$ XRD patterns were obtained using a (D5000 diffractometer) with a Cu$K\alpha$ line. In the glancing incidence geometry, the XRD analysis was performed in a Rigaku diffractometer also using a Cu$K\alpha$ radiation. The relative concentrations of carbon and nitrogen on the surface were obtained by X-ray photoelectron spectroscopy (XPS) using the 1 486.6 eV ($K\alpha$-line) photons from an Al target and a VG-CLAM2 hemispherical analyzer. The total apparatus resolution was ~0.85 eV (line-width plus analyzer). The system details are reported elsewhere.[13] The relative atomic composition at the sample surfaces was determined by integrating the core level peaks, properly weighted by the photoemission cross-section. As is well known, XPS gives information of the outermost atomic material layers (~0.5 nm).[14]

## Results and Discussion

Figure 1 shows the XRD pattern of nitrocarburized samples obtained at variable CH$_4$ content ($\phi$ = 0–4%) in the composition of the gaseous flux feeding the plasma chamber. The results from the pristine sample are also included. Due to the presence of carbon and nitrogen, the phases $\alpha$-Fe, $\gamma'$-Fe$_4$N, $\varepsilon$-Fe$_{2-3}$(C,N), and $\theta$-Fe$_3$C constitute the nitrocarburized layers. Roughly speaking, the intensity of XRD peaks of $\gamma'$-Fe$_4$N and $\varepsilon$-Fe$_{2-3}$(C,N) phases become progressively higher on CH$_4$ increasing in the gaseous mixture from $\phi$ = 0 to 3%. At higher proportions ($\phi$ = 4%), the $\gamma'$-Fe$_4$N and $\varepsilon$-Fe$_{2-3}$(C,N) phases disappear at the same time that the $\theta$-Fe$_3$C phase becomes more important. It is interesting to remark that at lower (1%) and higher (4%) $\phi$ the $\theta$-Fe$_3$C is present. From the diffraction patterns and taking into account the metallurgical properties, the condition $\phi$~3% for a pure $\varepsilon$-Fe$_{2-3}$(C,N) compound layer was obtained. Moreover, the GADRX of this sample shows a single phase of $\varepsilon$-Fe$_{2-3}$(C,N) in agreement with SEM result showing a compact and homogeneous compound layer up to 6 μm (Figure 2).

Figure 2 shows the SEM cross-section micrographs of the carbonitrided samples treated with $\phi$ = 0, 1, 2, 3, and 4% of CH$_4$ gaseous mixture feeding the chamber. In the samples treated with $\phi$ = 1, 2, and 3%, the electron microscopy revealed an increasing thickness of the carbonitrided layer. Also, there is a thin compound layer in the outermost surface followed by a diffusion zone containing carbon and nitrogen. For the sample prepared with $\phi$ = 4%, there is an abrupt change, i.e., no compound layer is formed and the diffusion layer is reduced. Also, different microstructure is observed as compared to samples treated at lower $\phi$. This reduction in the carbonitrided layer thickness is a result of the excessive carbon in the plasma atmosphere. This is probably due to the fact that excess of carbon reduces the nitrogen potential at the material surface. In other words, an excessive carbon content is accumulated on the surface material without having the time to diffuse deeper into the material. Moreover, the higher carbon content at the surface blocks the nitrogen penetration in the material.[3,9,15]

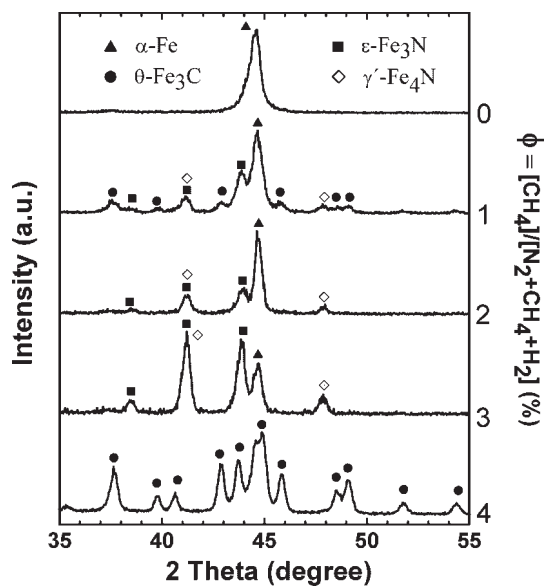

*Figure 1.* XRD pattern of the nitrocarburized samples as a function of the CH$_4$ content in the plasma atmosphere.

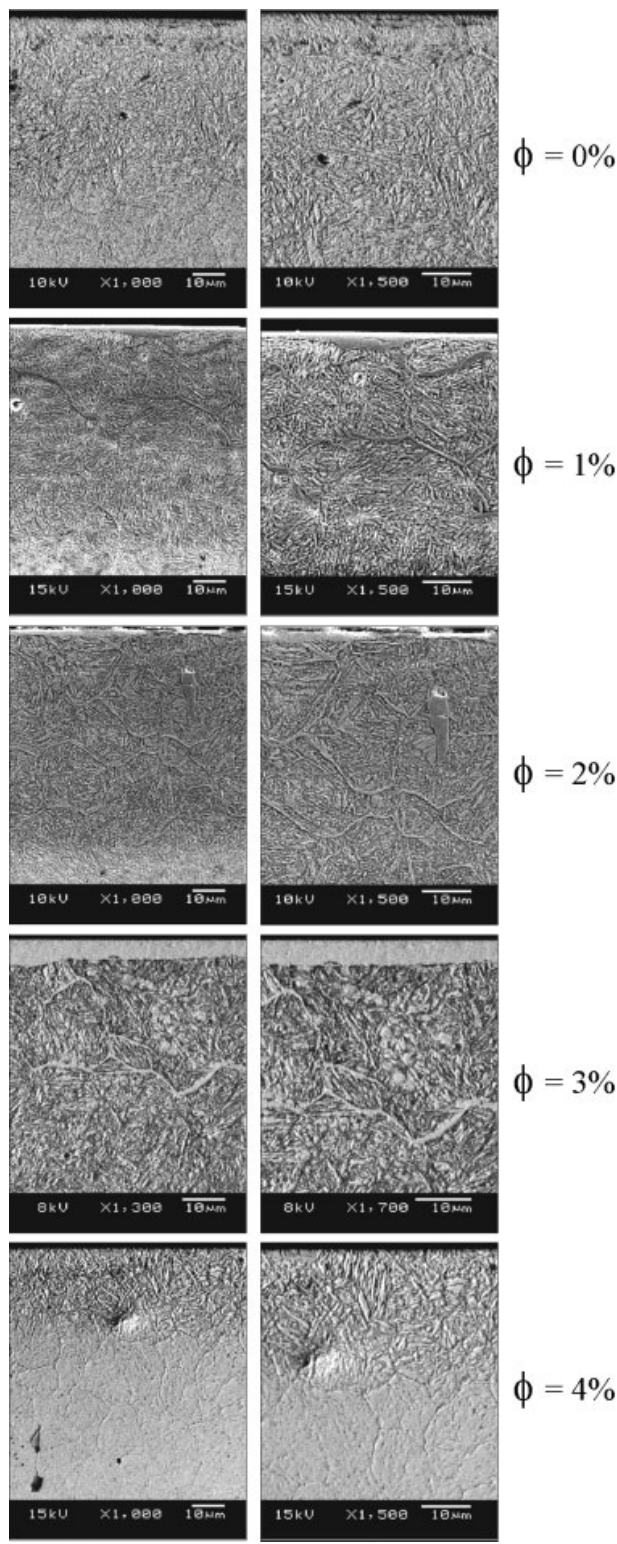

Figure 2. Cross-section SEM micrographs of the carbonitrided samples treated at different gaseous mixture ($\phi$) feeding the chamber.

As observed in Figure 3, a maximum hardness is obtained for $\phi = 3\%$. Indeed, the absence of $\theta$-Fe$_3$C seems to be the cause of this result (see Figure 1). Suhadi et al.[10] have shown that there is an optimal organic vapor concentration where the compound layer is maximum and there is no $\theta$-Fe$_3$C phase. These results suggest a complex correlation between the physical properties of the modified layers (bulk properties) and the CH$_4$ content in the plasma atmosphere (plasma properties). On the other hand, the hardness values are correlated with nitrogen and carbon concentration on the surface, as indicated by XPS. Indeed, carbon and nitrogen concentrations show a minimum and a maximum at $\phi = 3\%$, respectively. Moreover, the bulk layer properties show a simple relationship of the carbon and nitrogen concentrations measured by XPS in the material (Figure 4). This figure shows the hardness values at the surface samples as a function of the carbon and nitrogen concentration from XPS analysis. Harder layers are obtained at higher nitrogen and lower carbon contents, respectively. This condition is achieved for $\phi = 3\%$, where a thick, compact, homogeneous $\varepsilon$-Fe$_{2-3}$(C,N) layer was obtained with 2.6 wt.-% and 5.4 wt.-% of carbon and nitrogen respectively (Figure 2, 3). Finally, we note that, although the $\theta$-Fe$_3$C phase is hard, it must be avoided. This is so due to its poor mechanical properties and low toughness.[16] Last but not the least, we should note that the crystalline phases observed by XRD, GIXRD, and element concentrations (C and N) are in agreement with the Fe-C-N phase diagram proposed by Slycke et al.[9]

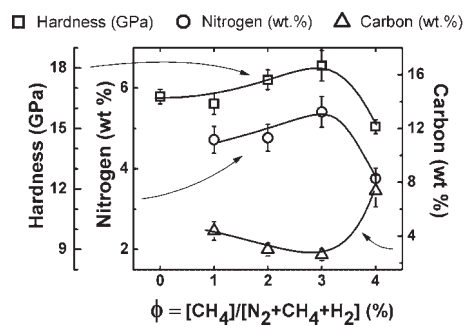

Figure 3. Hardness and nitrogen concentration (left axis) and carbon concentration (right axis) at 200 nm surface samples with different CH$_4$ content in the gaseous mixture feeding the chamber.

article published on Plasma Process. Polym. 2007, 4, S728–S731.
DOI: 10.1002/ppap.200731806

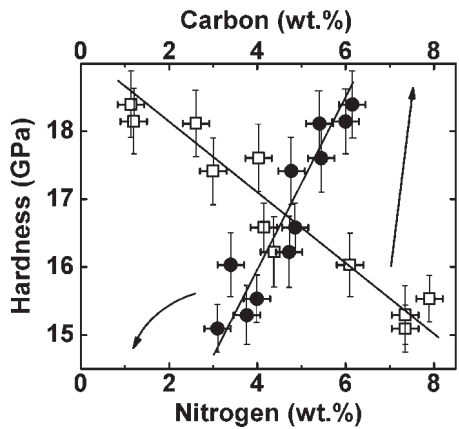

*Figure 4.* Hardness dependence at 200 nm depth (surface) as a function of carbon and nitrogen content.

## Conclusion

Samples of AISI H13 tool steel were treated by pulsed plasma nitrocarburizing process at variable $CH_4$ content in the plasma atmosphere. The samples were characterized by XRD, GIXRD, SEM, XPS, and hardness measurements. There is not a simple relationship between the gaseous mixture feeding the chamber (plasma structure) and the final physical properties of the modified layer (bulk properties). However, a fairly linear dependence among hardness, nitrided layer thickness, and N and C concentrations on material surface was observed. Increasing carbon content on the surface diminishes the compound and diffusion layers' thicknesses. A decreasing hardness due to the formation of $\theta$-$Fe_3C$ was also observed for increasing carbon content. These effects were ascribed to an excess carbon at the material surface blocking the nitrogen diffusion. A compact compound $\varepsilon$-$Fe_{2-3}$(C,N) monolayer was obtained with 2.6 wt.-% and 5.4 wt.-% of carbon and nitrogen concentration in the carbonitrided layer, respectively.

## References

[1] T. Bell, *Heat Treat. Met.* **1975**, *2(2),* 39.
[2] A. Wells, *J. Mater. Sci.* **1985**, *20,* 2439.
[3] T. Bell, Y. Sun, A. Suhadi, *Vacuum* **2000**, *59,* 14.
[4] M. Karakan, A. Alsaran, A. Çelik, *Mat. Design* **2004**, *25,* 349.
[5] A. Alsaran, M. Karakan, A. Çelik, F. Bulbul, I. Efeoglu, *J. Mater. Sci. Lett.* **2003**, *22,* 1759.
[6] K. Funatani, *Met. Sci. Heat Treat.* **2004**, *46,* 277.
[7] C. N. Chang, F. S. Chen, *Mater. Chem. Phys.* **2003**, *82,* 281.
[8] E. V. Pereloma, *Surf. Coat. Technol.* **2001**, *145,* 44.
[9] J. Slycke, L. Sproge, J. Agren, *Scand. J. Metall.* **1988**, *17(3),* 122.
[10] A. Suhadi, C. X. Li, T. Bell, *Surface* **2006**, *200,* 4397.
[11] L. F. Zagonel, C. A. Figueroa, R. DroppaJr., F. Alvarez, *Surf. Coat. Technoll* **2006**, *201,* 452.
[12] W. C. Oliver, G. M. Pharr, *J. Mater. Res.* **1992**, *7,* 1564.
[13] P. Hammer, N. M. Victoria, F. Alvarez, *J. Vac. Sci. Technol. A* **1998**, *16(5),* 2941.
[14] D. Briggs, M. P. Seah, Eds., ''Practical Surface Analysis, Auger and X-ray Photoelectron Spectroscopy'', John Wiley & Sons, New York 1990.
[15] Y. Sun, T. Bell, *Mater. Sci. Eng. A* **1991**, *140,* 421.
[16] R. W. K. Honeycombe, H. K. D H. Bhadeshia, ''Steel: Microstucture and Properties'', Edward Arnold, London 1995.